\documentclass[conference]{IEEEtran}
\IEEEoverridecommandlockouts
\usepackage{cite}
\usepackage{amsmath,amssymb,amsfonts}
\usepackage{algorithmic}
\usepackage{enumitem}
\usepackage{graphicx}
\usepackage{comment}
\usepackage{textcomp}
\usepackage{xcolor}
\usepackage{siunitx}
\def\BibTeX{{\rm B\kern-.05em{\sc i\kern-.025em b}\kern-.08em
    T\kern-.1667em\lower.7ex\hbox{E}\kern-.125emX}}

\usepackage[ruled,vlined]{algorithm2e}
\usepackage{algorithmic}

\begin{document}

\title{Grid-Connected, Data-Driven Inverter Control, \\ Theory to Hardware
}

\author{
    \IEEEauthorblockN{
        Sebastian Graf,\IEEEauthorrefmark{1} 
        Keith Moffat,\IEEEauthorrefmark{1} 
        Anurag Mohapatra,\IEEEauthorrefmark{3}
    }
    \IEEEauthorblockN{
        Alessandro Chiuso,\IEEEauthorrefmark{2}
        Florian D\"orfler\IEEEauthorrefmark{1}
    }
    \\
    \IEEEauthorblockA{\IEEEauthorrefmark{1}Automatic Control Laboratory, Swiss Federal Institute of Technology (ETH), Z\"urich, Switzerland \\ 
    Emails: grafse@student.ethz.ch, kmoffat@ethz.ch, dorfler@ethz.ch
    }
    \IEEEauthorblockA{\IEEEauthorrefmark{2}Dept. of Information Engineering, University of Padova, Italy \\ 
    Email: alessandro.chiuso@unipd.it}
    \IEEEauthorblockA{\IEEEauthorrefmark{3}Center for Combined Smart Energy Systems, Technical University of Munich, Munich, Germany \\ 
    Email: anurag.mohapatra@tum.de}
    \thanks{\noindent This research is supported by the Swiss National Science Foundation under the NCCR Automation (grant agreement 51NF40\_225155).}
}



\maketitle

\begin{abstract}
Grid-connected inverter control is challenging to implement due to the difficulty of obtaining and maintaining an accurate grid model. 
Direct Data-Driven Predictive Control provides a model-free alternative to traditional model-based control methods.
This paper describes how the recently-proposed Transient Predictive Control (TPC) can be used for real-world, plug-and-play inverter control. 
The following hypotheses were tested:
1) The TPC algorithm can be run online using standard hardware, and 
2) TPC, which is derived using Linear Time-Invariant assumptions, is effective for grid-connected inverter control, which is a nonlinear and time-varying system.
Experiments conducted on a two-converter benchtop setup and at the CoSES Laboratory on a 25 kVA converter connected to the Munich grid support these hypotheses. 
\end{abstract}

    

\section{Introduction}

Plug-and-play, grid-connected inverter control remains a challenge for the Power Electronics and Power Systems communities. 
As inverters have strict current limits, plug-and-play inverter control which proactively constrains output current is of particular interest.


The industry-standard grid-connected inverter control employs cascaded voltage and current control loops, tuned to ensure time-scale separation between the controllers \cite{ACMicrogrid_inverter}. 
To participate in power-balancing and voltage support, the current loops are often wrapped by additional droop-control loops or DC voltage control loops \cite{Teodorescu2010}.
Tuning the cascaded control loops requires repeated experiments and/or grid knowledge at the connection point including, for example, the short circuit and the $X/R$ ratios \cite{Teodorescu2010}. 
The real and reactive power droop control loops are decoupled based on this grid-connection knowledge and it is preferable for the resistance or the inductance to dominate the grid impedance to support this decoupling. 
However, in distribution grids the $X/R$ ratio is close to 1 and the droop decoupling assumptions are not well-supported.
Thus, the industry-standard methods are not plug-and-play and may not be reliable as we push for more active participation from grid-edge resources. 



Distribution grid parameters are generally not well known due to heterogeneity of components and the unknown dynamics of grid-connected prosumer devices \cite{challenges_dist_grid}.
Thus, plug-and-play inverter control methods generally require a data-driven aspect, such as parameter estimation \cite{Data-driven_survey}.
Additional methods include - 1. 
mode-estimation for a continuously updating state estimation of the grid \cite{probing2}, 2. sensitivity-based voltage and current control \cite{Gupta, Sensitivity_2}, and 3. neural-network-based approaches \cite{RL_control}.

``Direct'' Data-Driven Predictive Control (DDPC) \cite{coulson_data-enabled_2019, berberich2020data, dorfler2022bridging, breschi2023data, chiuso2025harnessing} provides an alternative data-driven approach which typically leverages Linear Time-Invariant (LTI) assumptions to construct controllers directly from data. Specifically, Data-Enabled Predictive Control (DeePC), proposed by \cite{coulson_data-enabled_2019}, has been applied to inverter control in \cite{huang_quadratic_2021} and \cite{huang_robust_2021}.
DeePC, however, requires a large optimization which may limit its real-world application.
Furthermore, DeePC is not causal and is biased when closed-loop training data are used \cite{moffat2025bias}.

This paper demonstrates the control of a grid-connected inverter with Transient Predictive Control (TPC) \cite{moffat_transient_2024}, demonstrating its feasibility for real-world application. 
TPC compresses the training data offline, producing a tractable online optimization problem.
The compression is done in such a way that the TPC prediction is causal, while also providing a consistent estimate (for LTI systems), regardless of whether the training data was gathered in open or closed-loop \cite{moffat_transient_2024}.

While the TPC theory is established for LTI systems, neither the grid nor the underlying control loops of inverter hardware are LTI. Furthermore, it is not a priori evident that 
the TPC optimization can be run fast enough to effectively control a grid-connected inverter.
Thus, the hypotheses tested in this paper are:
\begin{enumerate}
    \item TPC can be run online using standard hardware, and \label{hyp1}
    \item TPC 
    is effective for grid-connected inverter control, which is a nonlinear and time-varying system. \label{hyp2}
\end{enumerate}

To answer these hypotheses, two experiments were conducted. 
The first experiment was conducted on a lab bench using two inverters. One inverter was controlled with TPC, while the other inverter was controlled to behave as a grid-simulating infinite bus. This experiment provided a proof-of-concept and demonstrated that TPC can successfully limit the output current of the inverter. 
The first experiment provides an affirmative answer to Hypothesis (1) above.

The second experiment was conducted at the CoSES Laboratory at TU Munich \cite{CoSES_PHIL}, and used TPC to control a 25 kVA Egston COMPISO inverter connected to the Munich electric grid. 
To the authors' knowledge, this is the first real-world demonstration of grid-connected data-driven inverter control.
The second experiment provides an affirmative data point for Hypothesis (2) above.

The remainder of this paper is organized as follows. Section II provides the necessary background. Section III gives an overview of TPC inverter control. Sections IV and V describe the experiments and their results, and Section VI concludes the paper.




\section{Preliminaries}

\subsection{Notation}

Given a matrix A, its transpose is $A^T$. 
The $d$-dimensional identity matrix is $I_d$. 
The 2-norm weighted by a matrix $L$ is $\left\Vert \cdot \right\Vert_L$. 
For a sinusoidal signal such as $i$, the $dq$-components of the signal are $i_d$ and $i_q$ and its magnitude is $|i|$.
For any signal $w(t) \in \mathbb{R}^{n_w}$, we define the associated $\frac{1}{\sqrt{N}}$-scaled Hankel matrix with $N$ columns 
$W_{[t_0,t_1]} \in \mathbb{R}^{n_w(t_1-t_0+1) \times N}$ as:
\begin{align*}
    W_{[t_0,t_1]} := \frac{1}{\sqrt{N}}
    \left[\begin{smallmatrix}
        w(t_0) & w(t_0 + 1) & \hdots & w(t_0 + N - 1) \\
        w(t_0+1) & w(t_0 + 2) & \hdots & w(t_0 + N ) \\
        \vdots & \vdots & \ddots & \vdots \\
        w(t_1) & w(t_1 + 1) & \hdots & w(t_1 + N - 1) \\
    \end{smallmatrix}\right],
\end{align*}
where the $\frac{1}{\sqrt{N}}$ scaling normalizes the variance.


We consider the discrete-time, linear, and time-invariant (LTI) system class, whose input $u(t) \in \mathbb{R}^m$ and output $y(t) \in \mathbb{R}^q$ constitute a stationary joint process,
\begin{align*}
    z(t) = \begin{bmatrix} y(t) \\ u(t)  \end{bmatrix} \in \mathbb{R}^{q+m}.
\end{align*}

The future, $\tau$-long input and output sequences at time $t$ are
\begin{align*}
    y_f(t) &:= \begin{bmatrix} y^T(t+1) & \hdots & y^T(t + \tau) \end{bmatrix}^T \in \mathbb{R}^{q \tau} \text{ and} \\
    u_f(t) &:= \begin{bmatrix} u^T(t+1) & \hdots & u^T(t + \tau) \end{bmatrix}^T \in \mathbb{R}^{m \tau}.
\end{align*}
The future, $\tau$-long input and output references at time $t$ are $y_r(t)  \in \mathbb{R}^{q \tau}$ and $u_r(t)  \in \mathbb{R}^{m \tau}$, respectively.
The $\rho$-long lead-in measurement sequence at time $t$ is
\begin{align*}
    z_p(t) &:= \begin{bmatrix} z^T(t-\rho+1) & \hdots & z^T(t) \end{bmatrix}^T \in \mathbb{R}^{(q+m)\rho},
\end{align*}
which encodes the initial condition at time $t$.
For readability, we drop $(t)$ from the notation.

The certainty-equivalent multistep prediction \cite{moffat_transient_2024} is
\begin{align}\label{eqn:optMultiPred}
    \widehat{y}_f &= H \begin{bmatrix}
        z_p \\ u_f
    \end{bmatrix}:= \begin{bmatrix}
        H_p & H_u
    \end{bmatrix} \begin{bmatrix}
        z_p \\ u_f
    \end{bmatrix},  
\end{align}
where $H$ is the Multistep Predictor for $\tau$ steps into the future.

\subsection{Background}

\subsubsection{Grid-Connected Inverter Control}

\begin{figure}[h]
    \centering
    \includegraphics[width=.95\linewidth]{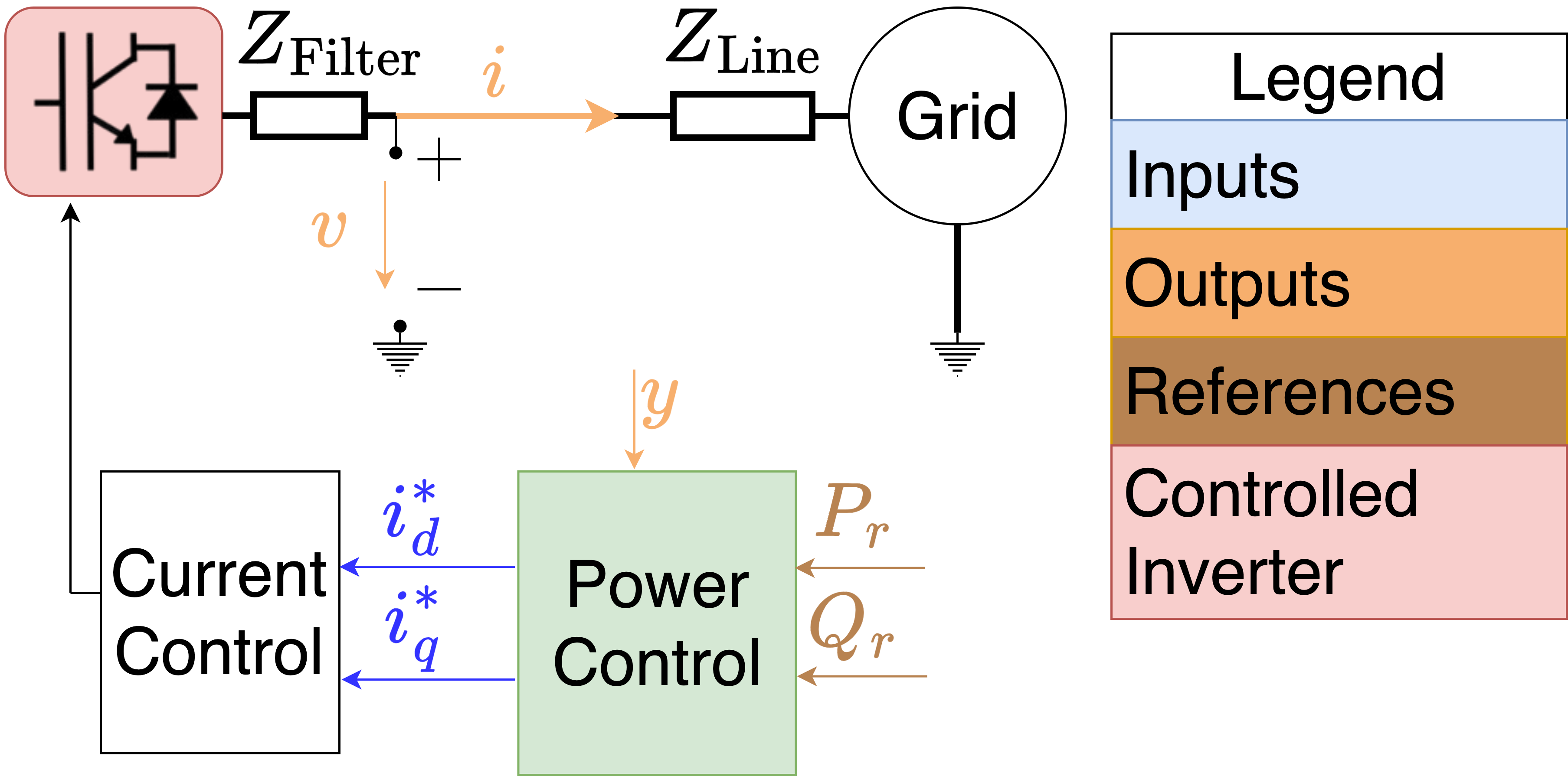}
    \caption{Grid-Connected Inverter Control}
    \label{fig:converter-setup}
\end{figure}

Figure \ref{fig:converter-setup} is the grid-connected inverter control circuit diagram that we focus on in this paper. 
The measurements of current $i$ and voltage $v$
are used to calculate real power $P$ and reactive power $Q$, which feed into the power control block through the output \( y \). 
$v$ and $i$ can be single, three-phase, or $dq$-reference frame measurements.
$y$ can be any combination of $v$, $i$, $P$, and $Q$, or pseudo-measurements constructed from those measurements, such as $|i|$, the output current magnitude.

The power controller determines the $dq$ reference frame current setpoints \( i_d^*, i_q^* \) such that $P$ and $Q$ track $P_r$, $Q_r$, the reference active and reactive power, respectively.
Thus, the ``input signal''  created by the power controller is
$
u = \begin{bmatrix}
    i_d^* & i_q^*
\end{bmatrix}^T,
$
which is passed to a standard current controller, which sends a voltage command to the inverter to track the current setpoints.

\subsubsection{Data-Driven Predictive Control Challenges}
In \cite{huang_quadratic_2021, huang_robust_2021}, the authors use DeePC to control power converters directly from data. 
DeePC solves the following optimization problem to obtain the optimal $u_f$ \cite{coulson_data-enabled_2019}:
\begin{equation}
    \begin{split}
        \min_{g, u_f, y_f} &\quad ||u_f-u_r||^2_{L_u} + ||y_f-y_r||^2_{L_y} + r_g(g) \\
s.t. &\quad 
\begin{bmatrix}
Z_{[1,\rho]} \\ U_{[\rho+1,\rho+\tau]}\\ Y_{[\rho+1,\rho+\tau]}
\end{bmatrix} 
g =
\begin{bmatrix} 
z_p \\ u_f \\ y_f
\end{bmatrix}, \\
 &\quad u_f \in \mathcal{U}, \ y_f \in \mathcal{Y}    \end{split}
    \label{eq:deepc}
\end{equation}
where $g \in \mathbb{R}^N$ 
and the sets $\mathcal{U} \subseteq \mathbb{R}^{m\tau}$ and 
$\mathcal{Y} \subseteq \mathbb{R}^{q\tau}$ 
are the feasible regions of the input $u$ and the output $y$, respectively.
$L_y \in \mathbb{R}^{q\tau \times q\tau}$ is the output cost matrix and 
$L_u \in \mathbb{R}^{m\tau \times m\tau}$ is the control cost matrix. 
$Z_{[1,\rho]},\ U_{[\rho+1, \rho+\tau]},\ Y_{[\rho+1, \rho+\tau]}$ are Hankel matrices constructed from input/output trajectories collected offline.
$r(g)$ is a regularization function
and, given $g^\star$, the argmin of \eqref{eq:deepc}, the optimal future input is 
$$u_f^\star=U_{[\rho+1, \rho+\tau]}g^\star.$$ 

As DeePC optimizes \( g \), which scales with \( N \), both the online computational load and memory usage depend on the number of training data points.
Furthermore, DeePC is a subspace-based method and thus does not assert causality \cite{moffat2025bias}.
DeePC has the following drawbacks:
\begin{itemize}
    \item it requires an large online optimization,
    \item the predictions are not causal \cite{moffat2025bias}, and
    \item it is biased when closed-loop training data are used \cite{moffat2025bias}.
\end{itemize}

\subsubsection{Transient Predictive Control}

TPC addresses these drawbacks by processing the training data (the input/output data from the system) offline.
Offline, TPC processes the training data to produce a consistent and causal estimate of the Multistep Predictor, $\widehat{H}$, by processing $Z_{[1,\rho+\tau]}$ with Algorithm 1 from \cite{moffat_transient_2024} (for more explanation, see \cite{moffat_transient_2024}):

\begin{algorithm}
    \SetKwInOut{Input}{input}
    \SetKwInOut{Output}{output}
    \caption{The Transient Predictor Method for estimating the Multistep Predictor} 
    \small{
    \Input{$Z_{[1,\rho+\tau]}$
    }
    \Output{$\widehat{H}$
    }
    \BlankLine
    
    $L \leftarrow$ \verb+LQ+$\left(Z\right)$

    $L^0$ $\leftarrow$ $L$ with the block-diagonal terms set to 0 
    
    $L^0_y$ $\leftarrow$ the $q\tau$ rows of $L^0$ corresponding to $y_f$ 
    
    $\widehat{\Phi}$ $\leftarrow$ $L^0_y L^{-1}$

    $\widehat{H}$ $\leftarrow$ $\begin{bmatrix}
    \left(I - \widehat{\Phi}_{y}\right)^{-1}\widehat{\Phi}_p & \left(I - \widehat{\Phi}_{y}\right)^{-1}\widehat{\Phi}_{u}
    \end{bmatrix}$
    }
    \label{alg:phiEst}
\end{algorithm}

Online, TPC solves the following optimization problem to determine the optimal $u_f$ at each timestep: 
\begin{align}
    \min_{u_f} & \quad \bigg\| \widehat{H} \begin{bmatrix}
        z_p \\ u_f
    \end{bmatrix} - y_r \bigg\|^2_{L_y} 
    + \big\| u_f - u_r \big\|^2_{L_u} + r(u_f, z_p),\nonumber \\ 
    \text{s.t.} & \quad u_f \in \mathcal{U},  \
                \widehat{H} \begin{bmatrix}
                   z_p \\ u_f
               \end{bmatrix} \in \mathcal{Y}, \label{eq:basic_opti}
\end{align}
where $r(u_f,z_p)$ is the optimal, quadratic regularization cost from \cite{chiuso2025harnessing}. 
In contrast to DeePC, TPC is causal and the online optimization does not include $g$ and therefore both the computational load and memory usage are independent of the number of training data points used.
\section{Proposed method}\label{sec:proposed_method}

The specific form of TPC inverter control that we implemented (other forms may produce different inputs than the $dq$ current references) solves \eqref{eq:basic_opti} to determine the inputs
\begin{align*}
    u = \begin{bmatrix}
        i_d^* & i_q^*
    \end{bmatrix}^T,
\end{align*}
which constitute the entries of $u_f$ for just the next timestep.
The reference angle used to convert between $dq$ components and timeseries measurements comes from an external source, such as a Phase-Locked Loop.

The output signal $y$ that is used to build the past trajectory $z_p$ and the future trajectory $y_f$ is a design choice and determines which quantities can be constrained/protected by \eqref{eq:basic_opti}.
That is, $\widehat{H}$ predicts the quantities that were included in the output data $y$ that was used to estimate $\widehat{H}$.

For the unconstrained case in which reference real and reactive power references are tracked, the output signal $y$ is $\begin{bmatrix}
    P & Q
\end{bmatrix}^T$. 
The next section describes how the output current can be constrained by including an additional output current signal in $y$, in addition to $P$ and $Q$, and two different options for the output current signal.

\subsection{Constraining Output Current}\label{sec:outputCurrentConstraint}

Inverters have strict output current limits determined by the hardware.
Thus, the output current magnitude $|i|$ of the inverter must be kept below the prescribed maximum $i_\text{max}$.
This can be done in two ways:
\begin{itemize}
    \item directly, by including \( |i| \) in \( y \), or
    \item indirectly, by including $i_d$ and $i_q$ in \( y \) and constraining $(i_d^2 + i_q^2)$. 
\end{itemize}
Directly including \( |i| \) in \( y \) results in \eqref{eq:basic_opti} being a Quadratic Program (QP). However, the relationship between the inputs $i_d^*$ and $i_q^*$ and \( |i| \) is nonlinear, which is challenging for LTI-based DDPC methods such as TPC.

On the other hand, the indirect output current constraint method, which instead includes $i_d$ and $i_q$ in \( y \) and constrains $(i_d^2 + i_q^2)$, results in \eqref{eq:basic_opti} being a Second-Order Conic Program (SOCP).
For this formulation, the relationship between the inputs $i_d^*$ and $i_q^*$ and the outputs $i_d$ and $i_q$ is closer to LTI.
Thus, the indirect SOCP method is preferable to the direct QP method, and we choose
\begin{align}\label{eqn:ydef}
    y = \begin{bmatrix}
        P & Q & i_d & i_q
    \end{bmatrix}^T.
\end{align}
No reference tracking cost is put on the $i_d$ and $i_q$ elements of $y$.
\section{Two-Converter Experiment}

The two-converter experiment tests Hypothesis (\ref{hyp1})---
\textit{TPC can be run online using standard hardware.}

\subsection{Experiment set-up}

\begin{figure}[h]
    \centering
    \includegraphics[width=.95\linewidth]{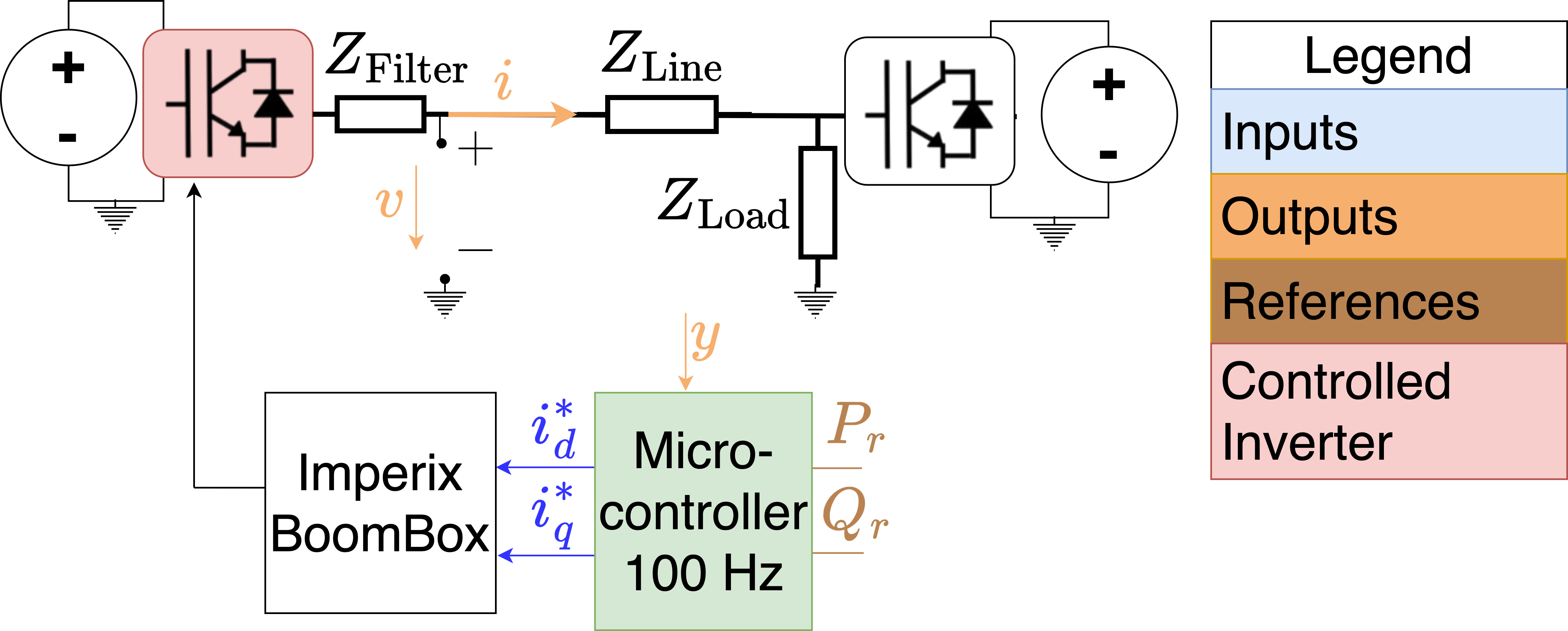}
    \caption{Schematic of the two-converter experiments}
    \label{fig:phys-set-up}
\end{figure}

Figure \ref{fig:phys-set-up} describes the hardware setup, which consists of two 3 kVA back-to-back DC-AC converters: one inverter acting as an Infinite Bus (\(v_d=1\), \(v_q=0\), \(f=50\) Hz), and the other inverter controlled by TPC to track a power setpoint. 
The controller hardware is the Imperix BoomBox, which manages real-time control of IGBT bridges at 8 kHz, alongside a microcontroller (model STM32H723 with a 32-bit Arm Cortex M7 Core and 564 Kbytes RAM, costing $\sim$\$10) executing TPC at 100 Hz using the ECOS solver. 


Prior to deployment, training data is collected by applying white-noise $i_d^*$ and $i_q^*$ inputs to the system and building the Hankel matrix $Z_{[1,\rho+\tau]}$ with 500 data point-long input and output trajectories.
$Z_{[1,\rho+\tau]}$ is passed to Algorithm 1 to produce $\widehat{H}$, which is computed offline and then loaded onto the microcontroller's RAM. The TPC parameters are $\tau=\rho=6$, 
$L_y=\text{diag}(4.5e5,4.5e5,0,0) \otimes I_\tau$, and $L_u=\text{diag}(0.001, 0.001) \otimes I_\tau$.

During operation, the microcontroller receives the measurements \( y \) \eqref{eqn:ydef} from the BoomBox 
at each timestep and computes \( i_d^* \) and \( i_q^* \) by solving \eqref{eq:basic_opti} and taking the input for just the next timestep.
The microcontroller then sends \( i_d^* \) and \( i_q^* \) to the BoomBox, which adjusts the converter's PWM signals accordingly.




\subsection*{Memory demands and computational load}


Using the ECOS solver on the microcontroller, a full-sized Hankel matrix with 500 data points ($N_\text{data} = 500$) causes DeePC to exceed the microcontroller's memory limit of 564 Kbytes. After reducing the number of data points to 50 so that the Hankel matrix fits on the microcontroller, DeePC's maximum speed is 10 Hz, which does not meet the 100 Hz specification.
TPC, on the other hand, does not encounter problems running at 100 Hz on the microcontroller.

To test the impact of computational power, we also experimented with DeePC and TPC on a laptop computer.
TPC's solve times remain fast (70 µs), regardless of the number of data used to estimate the Multistep Predictor with Algorithm \ref{alg:phiEst}, as expected. 
DeePC's solve time increases with the number of data, which increases the size of the Hankel matrix, limiting its performance; with 500 data points, it runs at just 25 Hz.


\begin{table}[h]
    \centering
    \begin{tabular}{l|c|c}
         $N_{\text{data}}$ &  \textbf{DeePC} & \textbf{TPC} \\
         \hline
         \multicolumn{3}{c}{Micro-controller using ECOS solver}\\
         \hline
         50 &  100 ms  & 3 ms\\
         500 &  -  & 3 ms\\
         \hline
         \multicolumn{3}{c}{Laptop Computer 
         }\\
         \hline
         100 &  2 ms & 70 µs\\
         500 &  40 ms & 70 µs\\
    \end{tabular}
    \vspace{2ex}
    \caption{Solve time comparison between DeePC and TPC}
    \label{tab:comp_speed}
\end{table}

\subsection{Experiments}

We tested the TPC-controlled converter's response to a step change in the active power reference \( P_r \) from 0 to 0.3 p.u., 
with \( Q_r \) fixed at 0 p.u.. 
This test simulates TPC's response to a sudden load increase. 
Two experiments were conducted---one without constraints and one in which a current magnitude constraint of 0.2 p.u. limits the output current.

\begin{figure}[h]
    \centering
    \includegraphics[width=\linewidth]{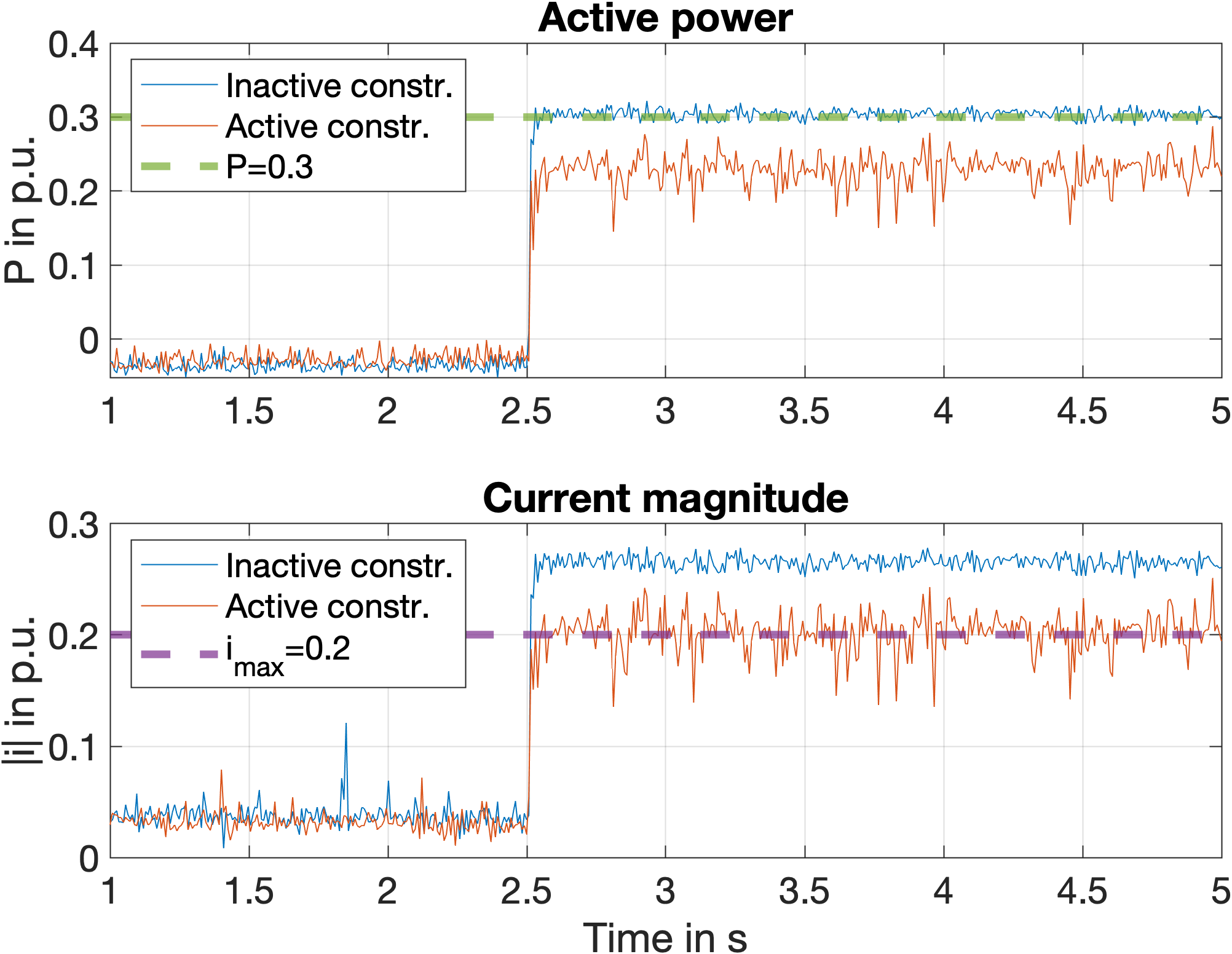}
    \caption{Performance of TPC in the two-converter experiment when $P_r$ steps from 0 to 0.3 p.u. with active and inactive current constraints}
    \label{fig:constrained_TPC}
\end{figure}

When the current magnitude constraint is inactive, 
TPC effectively tracks the step response. 
When the constraint is active, TPC enforces the current limitation, which results in reduced output power, as expected. 
The experiments are plotted in Figure \ref{fig:constrained_TPC}.

\subsection{Lessons from the two-converter experiments}


The computation benchmark in Table \ref{tab:comp_speed} and the demonstration of the desired step responses in Figure \ref{fig:constrained_TPC} confirm Hypothesis \ref{hyp1}.
Furthermore, the TPC-controlled inverter is able to track power setpoints when connected to an approximately-constant voltage source, and the output constraint formulation described in section \ref{sec:outputCurrentConstraint} prevents the inverter from violating the inverter output current magnitude constraint.

\section{Grid-Connected Inverter Experiments}

The grid-connected inverter experiments test Hypothesis (\ref{hyp2})---\textit{TPC 
is effective for controlling a grid-connected inverter, which is a nonlinear, time-varying system.}

\subsection{Experiment set-up}

\begin{figure}[h]
    \centering
    \includegraphics[width=\linewidth]{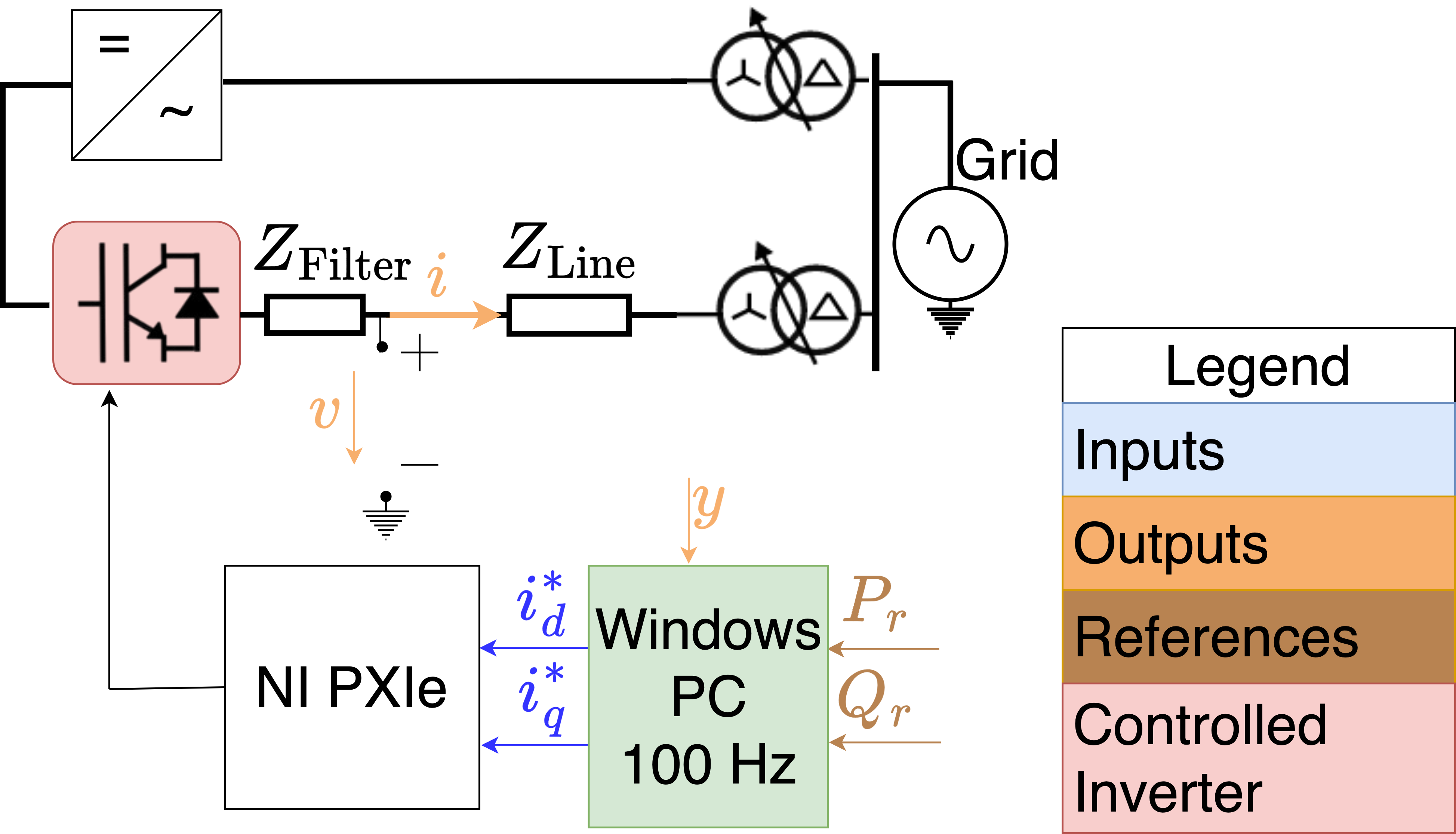}
    \caption{Schematic of the experiments at TUM's CoSES Lab \cite{CoSES_PHIL}}
    \label{fig:tum_schematic}
\end{figure}

Figure \ref{fig:tum_schematic} depicts the hardware setup, which consists of a four-leg Egston COMPISO inverter with a base power of 25 kVA. The inverter is connected via a 70 sqmm power cable to the Munich distribution grid through a 250 kVA MV/LV transformer. The DC current comes from a 4-quadrant rectifier.  
The inverter current control loop operates at 5 kHz in real-time on an NI PXIe 8880 RT controller, while the TPC operates at 100 Hz on a Windows PC. A detailed description of the electrical and control setup at CoSES can be found in \cite{CoSES_PHIL}. The same TPC parameters are used as in the previous section.

\subsection{Experiments}

\begin{figure}[h]
    \centering
    \includegraphics[width=\linewidth]{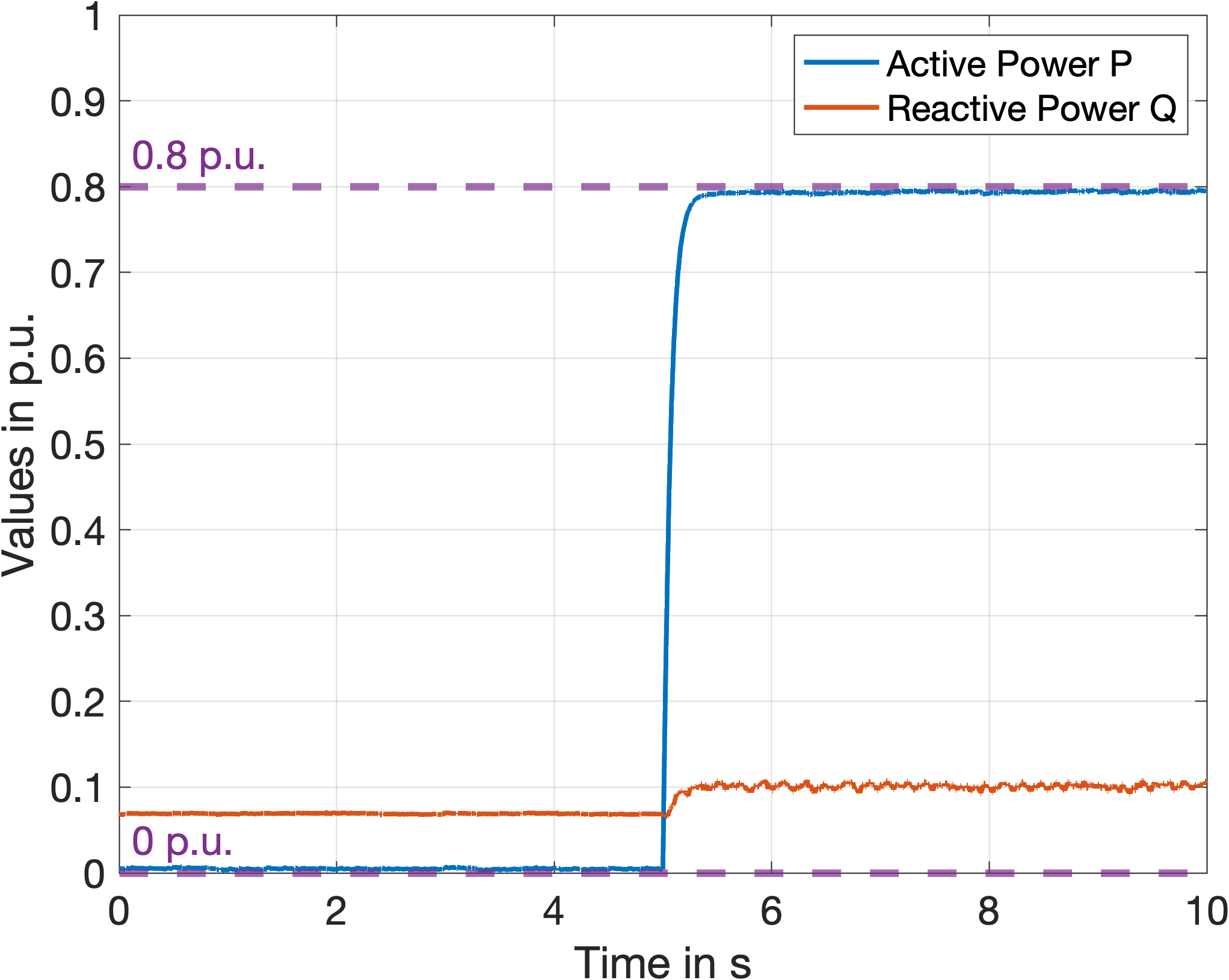}
    \caption{Step Response of the TPC on the TUM system with $P_r:$ 0 $\rightarrow$ 0.8 p.u.; $Q_r$ = 0.1 p.u.. }
    \label{fig:F-->B}
\end{figure}

\begin{figure}[h]
    \centering
    \includegraphics[width=\linewidth]{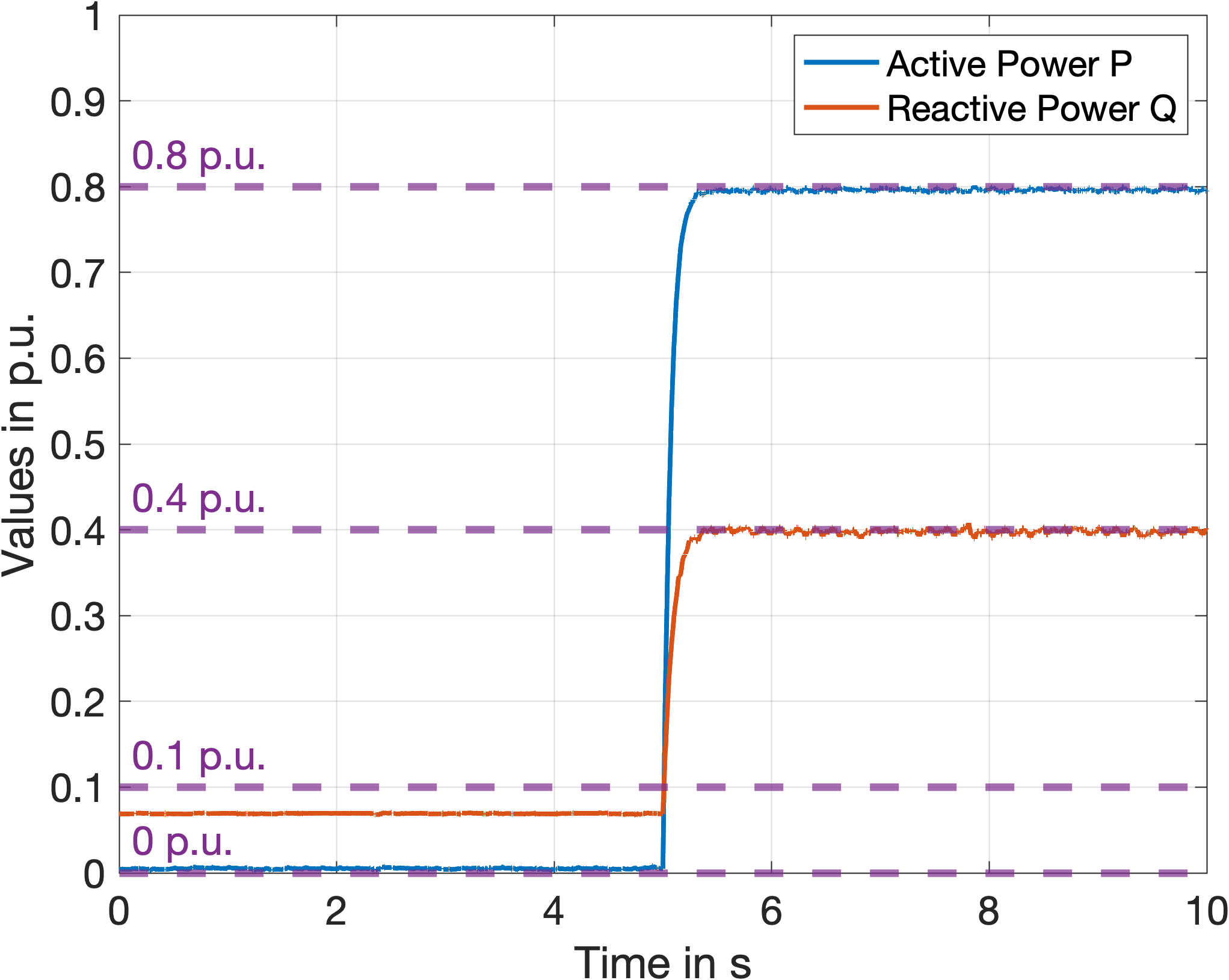}
    \caption{Step Response of the TPC on the TUM system with $P_r:$ 0 $\rightarrow$ 0.8 p.u.; $Q_r$ = 0.1 p.u. $\rightarrow$ 0.4 p.u..}
    \label{fig:F-->D}
\end{figure}

Figures \ref{fig:F-->B} and \ref{fig:F-->D} each plot the active and reactive power step-responses for the grid-connected, TPC-controlled inverter.
In Figure \ref{fig:F-->B}, we step-change the active power reference from 0 to 0.8 p.u. while keeping the reactive power constant at 0.1 p.u.. 
This case is similar to the operation of a photovoltaic inverter feeding a desired amount of active power to the grid. 
Figure \ref{fig:F-->B} shows that the controller is able to track the P setpoint without overshoot. 
The Q setpoint is slightly below the 0.1 p.u. and there is a small perturbation on the injected Q at the step-change. 
These tracking errors can be attributed to the non-linearities of the switching amplifier at low reactive power setpoints.

In Figure \ref{fig:F-->D}, we mimic an inverter performing reactive power support along with active power injection, an important grid ancillary service for voltage control. 
We step-change the active power reference from 0 to 0.8 p.u., and the reactive power reference from 0.1 to 0.4 p.u.. 
The inverter is able to track the P and Q setpoints (though a slight mismatch in Q at the low setpoint persists). 

\subsection{Lessons from the grid-connected inverter experiments}

The demonstrations of the desired step-responses of a grid-connected inverter in Figures \ref{fig:F-->B} and \ref{fig:F-->D} provide a positive data point for Hypothesis \ref{hyp2}. 
Elaborating, while the grid is not LTI, the Munich grid from the perspective of the Egston inverter at CoSES was ``LTI-enough'' for TPC to be effective for the specified experiments.
Fully confirming Hypothesis \ref{hyp2} requires testing TPC on larger inverters and in more grid-connected scenarios, e.g., with different grid dynamics, line impedances, and phase balances.

\section{Conclusion}

This paper describes how TPC can be used for real-world, plug-and-play inverter control.
The two-converter experiments demonstrated that TPC's online optimization is not prohibitive.
The grid-connected experiments demonstrated that TPC was effective for the tests conducted. 
Future work includes testing TPC inverter control in more grid-connected scenarios such as unbalanced three-phase and fault scenarios, and testing TPC on larger inverters.






\addcontentsline{toc}{chapter}{Bibliography}

\bibliographystyle{IEEEtran}
\bibliography{biblio}

\appendices

\end{document}